RESEARCH ARTICLE

# HTC Vive MeVisLab integration via OpenVR for medical applications

Jan Egger[1,2]*, Markus Gall[1], Jürgen Wallner[3], Pedro Boechat[3], Alexander Hann[4], Xing Li[5], Xiaojun Chen[5], Dieter Schmalstieg[1]

1 Institute of Computer Graphics and Vision, Graz University of Technology, Inffeldgasse 16c/II, Graz, Austria, 2 BioTechMed-Graz, Krenngasse 37/1, Graz, Austria, 3 Medical University of Graz, Department of Oral and Maxillofacial Surgery, Auenbruggerplatz 5/1, Graz, Austria, 4 Department of Internal Medicine I, Ulm University, Albert-Einstein-Allee 23, Ulm, Germany, 5 Shanghai Jiao Tong University, School of Mechanical Engineering, Shanghai, China

* egger@tugraz.at

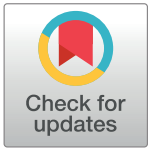

## Abstract

Virtual Reality, an immersive technology that replicates an environment via computer-simulated reality, gets a lot of attention in the entertainment industry. However, VR has also great potential in other areas, like the medical domain, Examples are intervention planning, training and simulation. This is especially of use in medical operations, where an aesthetic outcome is important, like for facial surgeries. Alas, importing medical data into Virtual Reality devices is not necessarily trivial, in particular, when a direct connection to a proprietary application is desired. Moreover, most researcher do not build their medical applications from scratch, but rather leverage platforms like MeVisLab, MITK, OsiriX or 3D Slicer. These platforms have in common that they use libraries like ITK and VTK, and provide a convenient graphical interface. However, ITK and VTK do not support Virtual Reality directly. In this study, the usage of a Virtual Reality device for medical data under the MeVisLab platform is presented. The OpenVR library is integrated into the MeVisLab platform, allowing a direct and uncomplicated usage of the head mounted display HTC Vive inside the MeVisLab platform. Medical data coming from other MeVisLab modules can directly be connected per drag-and-drop to the Virtual Reality module, rendering the data inside the HTC Vive for immersive virtual reality inspection.

**Data Availability Statement:** Data is available on figshare at https://figshare.com/articles/Cranial_Defect_Datasets/4659565.

**Funding:** The work received funding from BioTechMed-Graz in Austria (https://biotechmedgraz.at/en/, "Hardware accelerated intelligent medical imaging"), the 6th Call of the Initial Funding Program from the Research & Technology House (F&T-Haus) at the Graz University of Technology (https://www.tugraz.at/en/, "Interactive Planning and Reconstruction of Facial Defects", PI: Jan Egger) and was supported

## Introduction

Virtual Reality (VR) places a user inside a computer-generated environment. This paradigm is becoming increasingly popular, due to the fact that computer graphics have progressed to a point where the images are often indistinguishable from the real world. The computer-generated images previously presented in movies, games and other media are now detached from the physical surroundings and presented in new immersive head-mounted displays (HMD), like the Oculus Rift, the PlayStation VR or the HTC Vive (Fig 1). Virtual reality not only places a user inside a computer-generated environment – it is able to completely immerse a user in a virtual world, removing any restrictions on what a user can do or experience [1–3]. Beside movies, games, and other media, the medical area has great potential for the newly VR devices, by TU Graz Open Access Publishing Fund. Dr. Xiaojun Chen receives support by the Natural Science Foundation of China (www.nsfc.gov.cn, Grant No.: 81511130089) and the Foundation of Science and Technology Commission of Shanghai Municipality (Grants No.: 14441901002, 15510722200 and 16441908400). The funders had no role in study design, data collection and analysis, decision to publish, or preparation of the manuscript.

**Competing interests:** The authors declare no competing interests.

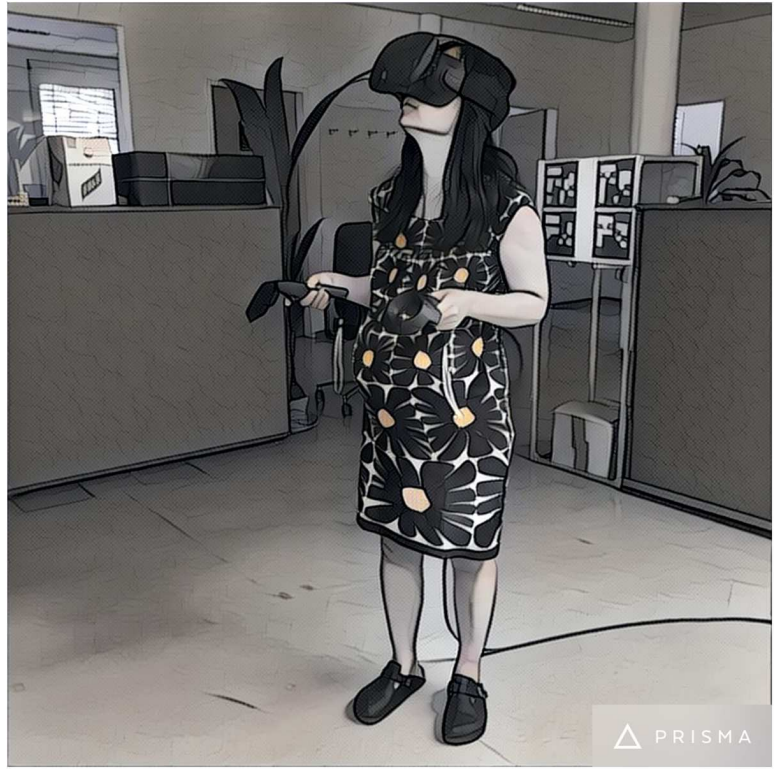

**Fig 1. Illustration of a person using the HTC Vive and controllers (photo filtered with PRISMA 2.2.1 under an iPhone 6s running iOS 9.3.3).**

https://doi.org/10.1371/journal.pone.0173972.g001

because they are also now able to process and display high-resolution medical image data acquired with modern CT (Computed Tomography) and MRI (Magnetic Resonance Imaging) scanners [4]. Examples are surgery trainers and simulators, and preoperative surgical planning [5], [6]. Others already working in the area of medical Virtual Reality are Nunnerley et al. [7], who tested the feasibility of an immersive 3D virtual reality wheelchair training tool for people with spinal cord injury. They designed a wheelchair training system that used the Oculus Rift headset and a joystick. Newbutt et al. [8] studied the usage of the Oculus Rift in autism patients. Their study explores the acceptance, willingness, sense of presence and immersion of participants with autism spectrum disorder (ASD). The examination of digital pathology slides with the virtual reality technology and the Oculus Rift has been explored by Farahani et al. [9]. They applied the Oculus Rift and a Virtual Desktop software to review lymph node cases for digital pathology. A usability comparison between HMD (Oculus Rift) and stereoscopic desktop displays (DeepStream3D) in a VR environment with pain patients has been performed by Tong et al. [10]. Twenty chronic pain patients assessed the severity of physical discomforts by trying both displays, while watching a virtual reality pain management demo in a clinical setting. An interactive 3D virtual anatomy puzzle for learning and simulation has been designed and tested by Messier and collages [11]. The virtual anatomy puzzle is supposed to help users to learn the anatomy of various organs and systems by manipulating virtual 3D data. It was implemented with an Oculus Rift. A computer-based system, which can record a surgical procedure with multiple depth cameras and reconstruct in three dimensions the dynamic geometry of the actions and events that occur during the procedure, has been introduced by Cha et al. [12]. Equipped with a virtual reality headset, such as the Oculus Rift, the user was able to walk around the reconstruction of the procedure room, while controlling the playback of the recorded surgical procedure with simple VCR controls (e.g., play, pause, rewind, and fast forward). Xu et al. [13] studied the accuracy of the Oculus Rift during cervical spine mobility measurement by designing a virtual reality environment to guide participants to perform certain neck movements. Subsequently, the cervical spine kinematics was measured by both the Oculus Rift tracking system and a reference motion tracker. The Oculus Rift has been applied by Kim et al. [14] as a cost-effective tool for studying visual-vestibular interactions in self-motion perception. The vection strength has been measured in three conditions of visual compensation for head movement (1. compensated, 2. uncompensated and 3. inversely compensated). King [15] developed an immersive VR environment for diagnostic imaging. The environment consisted of a web server acquiring data from volumes loaded within the 3D Slicer platform [16] and forwarding them to a Unity application (https://unity3d.com/) to render the scene for the Oculus Rift. However, to the best of the authors' knowledge, no work described the integration the HTC Vive into MeVisLab (http://www.mevislab.de/) platform [17] yet. We developed and implemented a new module for the medical prototyping platform MeVisLab that provides an interface via the OpenVR library to head mounted displays, enabling the direct and uncomplicated usage of the HTC Vive in medical applications. Unlike the Oculus Rift, the room-scale tracking offered by the HTC Vive allows walking around virtual objects, which enables a more advanced immersion and inspection.

This contribution is organized as follows: Section 2 introduces details of the methods, Section 3 presents experimental results and Section 4 concludes the paper and gives an outlook on future work.

## Methods

In this section, the materials and methods that have been used for the integration of the HTC Vive into the MeVisLab environment via the OpenVR library are introduced.

### Datasets

For testing and evaluating our software integration, we used multiple high-resolution CT (Computed Tomography) scans from the clinical routine. The resolution of the scans consisted of 512x512 voxels in the x- and the y-direction with an additional few hundred slices in the z-direction. The scans varied in anatomy and location of pathology, including datasets from patient skulls with cranial defects. In Fig 2, 3D visualizations of patient skulls with cranial defects on the left (left) and right side (right) are shown. The medical scans are freely available for download and usage in own research projects, however, we kindly ask users cite our work [18], [19]. All relevant data are hosted at the public repository Figshare. Please see data hosted at Figshare at the following URL:

https://figshare.com/articles/Cranial_Defect_Datasets/4659565.

Note that the datasets from our clinical partners have not been altered or downsampled in any way. Hence, we assess the visual quality and evaluate the frames per second (fps) when displaying original sized scans inside the HTC Vive.

### MeVisLab

This paragraph describes the medical imaging platform MeVisLab (Medical Visualization Laboratory), which includes a software development kit (SDK) used to interface with the HTC Vive. The MeVisLab platform provides basic and advanced algorithms for medical image processing and visualization. It also includes an environment for programming new modules in C++, as well as an environment for implementing user interfaces with MDL script (MeVis Description

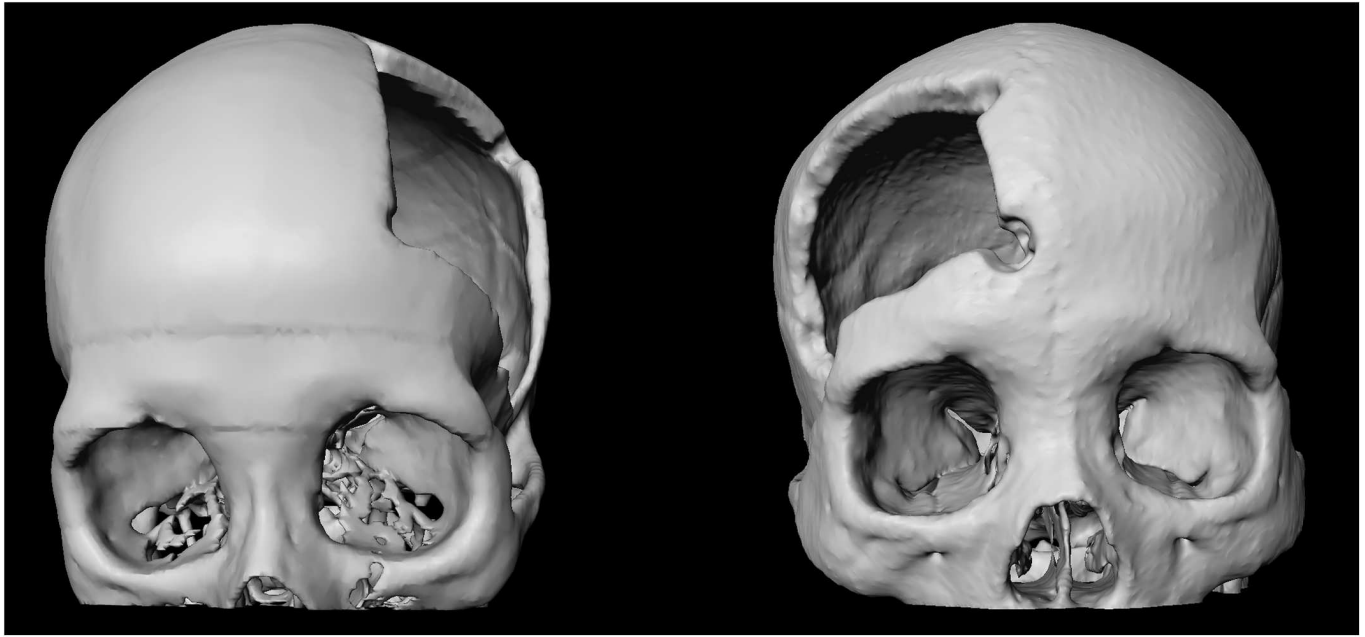

**Fig 2. 3D visualization of a patient skull with a cranial defects on the left side (left) and 3D visualization of a patient skull with a cranial defects on the right side (right).**

https://doi.org/10.1371/journal.pone.0173972.g002

Language). MeVisLab allows to construct dataflow networks using a variety of image processing modules. Developers can employ visual programming by compositing pre-installed modules into complex networks using the graphical user interface. Alternatively, the Python scripting language can be used. Finally, C++ programmers generate individually designed modules. Furthermore, it is possible to design user interfaces with MDL script, hiding the complexity of user-interface programming from non-technical users. Fig 3 shows a simple MeVisLab network with a small amount of modules: a module for loading the image data, an algorithm module, which applies a threshold function in this example, and a viewer module for the visualization.

In summary, the MeVisLab network concept is based on a graphical representation of modules with specific functions for image processing. MeVisLab uses three different types of modules (Fig 4). ML modules (blue) are processing modules, Open Inventor modules (green) provide scene graphs in 3D, and Macro modules (brown) combine other modules. Module connectors located on the bottom of the module are inputs, and connectors located at the top of the modules are outputs. The shape of the inputs and outputs define the connection types: A triangle for ML images, a rectangle for a base object indicating pointers to data structures, and a half circle for an Open Inventor scene (see Figs 3 and 4). A data transmission between connectors is enabled by a connection, visually represented by a blue line, like shown in Fig 3 on the right side. In addition to these data connections, a parameter connection can also be established. This enables the connection of single parameters between modules and is indicated by a two-sided arrow. For further information, please see the MeVisLab Reference manual (last accessed in January 2017):

http://mevislabdownloads.mevis.de/docs/current/MeVisLab/Resources/Documentation/Publish/SDK/MeVisLabManual/index.html.

## HTC Vive

This paragraph states the technical specifications of the VR headset HTC Vive, which is developed by HTC and Valve Corporation and was released in April 2016. In contrast to the current

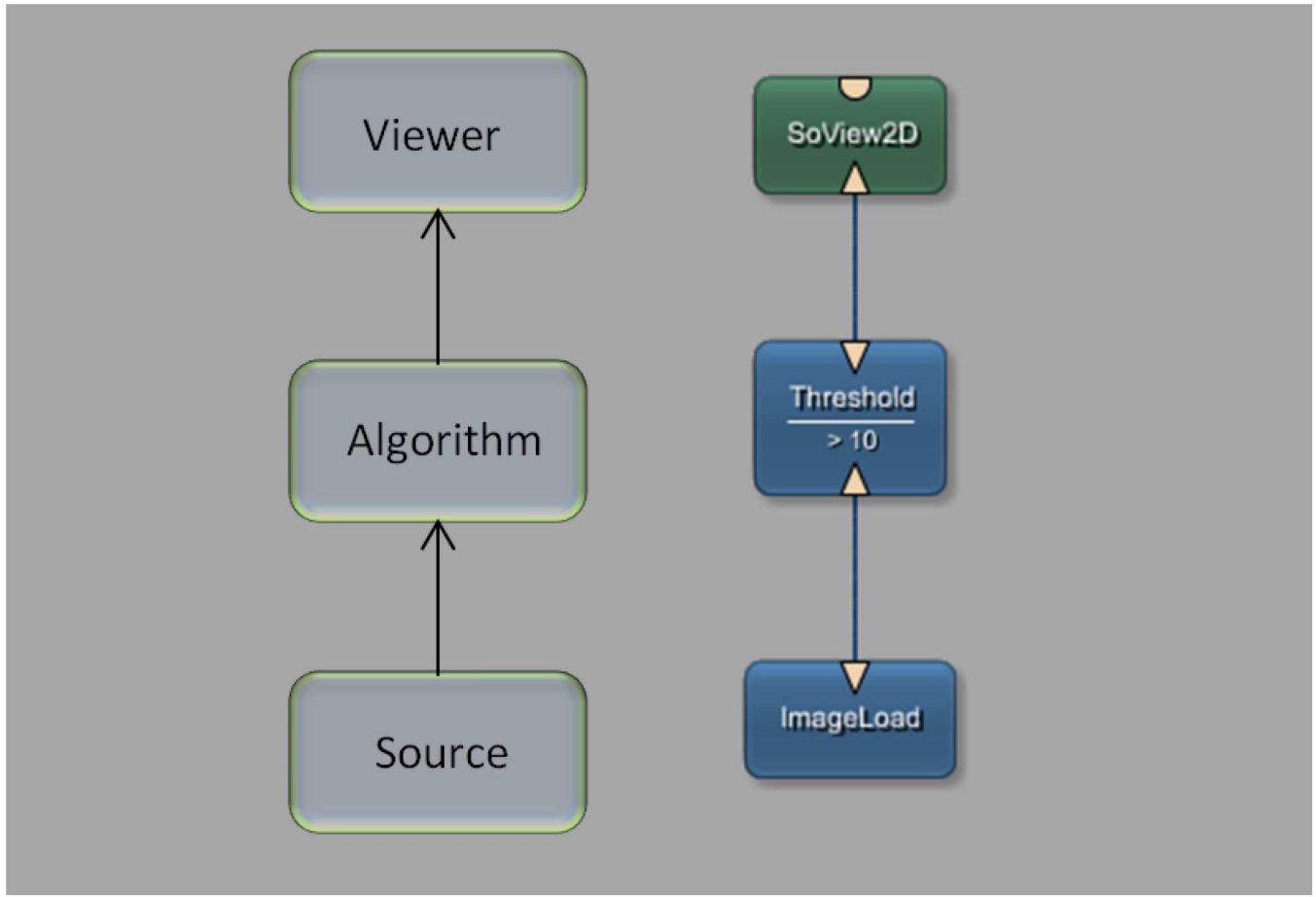

**Fig 3. Basic module processing pipeline with a network of three modules: An image source for loading the data set into the network, an algorithm module to process the data and a viewer module that enables the visualization of the image.**

https://doi.org/10.1371/journal.pone.0173972.g003

version of the Oculus Rift, the HTC Vive is designed to turn a room into 3D space. Two stationary reference units track the user's head and handheld controllers, while the user is free to walk around and manipulate virtual objects. The HTC Vive uses an organic light-emitting diode (OLED) display and provides a combined resolution of 2160x1200 (1080x1200 per eye) with a refresh rate of 90 Hz and a field of view (FOV) of about 110 degrees. The head mounted display weights about 555 grams and has HDMI 1.4, DisplayPort 1.2 and USB 2.0 connections.

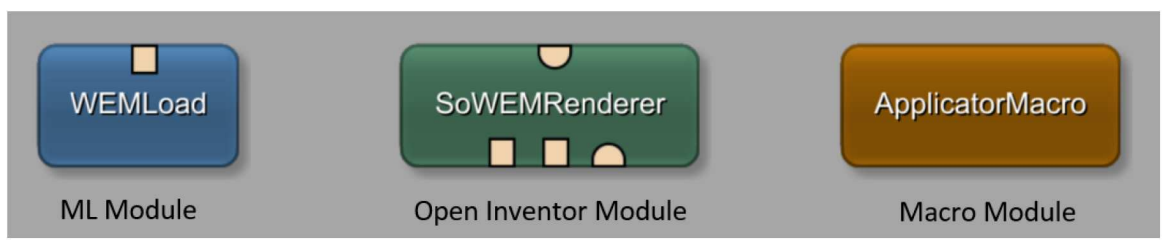

**Fig 4. Types of modules under MeVisLab: ML modules (blue) are page-/patch based and demand-driven, Open Inventor modules (green) provide scene graphs in 3D and Macro modules (brown) consist of a combination of modules.**

https://doi.org/10.1371/journal.pone.0173972.g004

Moreover, the HTC Vive has a 3.5 mm audio jack for headphones and a built-in microphone. Finally, the HTC Vive has a front-facing camera for blending real-world elements into the virtual world. For further information, please see the HTC Vive website (last accessed in January 2017):

https://www.htcvive.com

### OpenVR

This paragraph gives a short description of OpenVR, which is a software development kit and application programming interface developed by Valve. The C++ library supports the HTC Vive and other VR headsets via the SteamVR software platform (http://steamvr.com). Thus, the OpenVR API provides a way to connect and interact with VR displays without relying on a specific hardware vendor's SDK. OpenVR is implemented as a set of C++ interface classes with virtual functions supposed to support also future versions of the hardware. For further information, please see the corresponding GitHub repository (last accessed in January 2017):

https://github.com/ValveSoftware/openvr

The current OpenVR GitHub repository comes for Microsoft Windows, Linux and MacOS. Moreover, it includes examples in C++ for Visual Studio from Microsoft.

### Workflow

Fig 5 shows a high-level workflow diagram of the communication and interaction between MeVisLab and the HTC Vive device via OpenVR. In doing so, the MeVisLab platform provides several modules to import and load medical data, e.g., in DICOM format (http://dicom.nema.org/). The medical image data can be processed and visualized with a variety of modules under MeVisLab. Options include the visualization in 2D and 3D, or medical image analysis algorithms, like the segmentation of anatomical or pathological structures. Moreover, MeVisLab modules that derive directly from ITK (Insight Segmentation and Registration Toolkit, https://itk.org/) and VTK (The Visualization Toolkit, http://www.vtk.org/) can be applied.

Our new HTCVive module communicates with the HTC Vive via OpenVR. However, our MeVisLab module may also be able to communicate with another VR device, like the Oculus Rift (https://www.oculus.com/), because the OpenVR API provides a way to connect and interact with VR displays without relying on a specific hardware vendor's SDK. The HTCVive module transfers the image date to the HMD and receives the position and the orientation of the HMD in real-world to MeVisLab. The position and orientation enable further visualizations under MeVisLab, like rendering the user's view in a second 3D viewer on a conventional screen, so users without HMD can see the VR view, too. For our implementation, we followed the hellovr_opengl source code example from OpenVR (https://github.com/ValveSoftware/openvr/tree/master/samples/hellovr_opengl). In doing so, using an initialization method to start the SteamVR runtime and render a distortion view of the right and left eye.

### Network

Fig 6 shows the overall MeVisLab network with our HTCVive module (blue) in the lower left surrounded by a yellow rectangle. In this example network, the (medical) data is imported via the WEMLoad module (here named DataLoad) and directly passed via its output (rectangle) to the HTCVive module (rectangle input at the bottom of the HTC Vive module). Furthermore, the loaded data is passed via a WEMModify module (blue, lower right side), an SoWEMRenderer module (green, right side in the middle) and an SoSeparator module (green, upper right side) to another SoSeparator module (named 3DUserView at the top). The window on the right side belongs to the 3DUserView module at the top and displays what the user

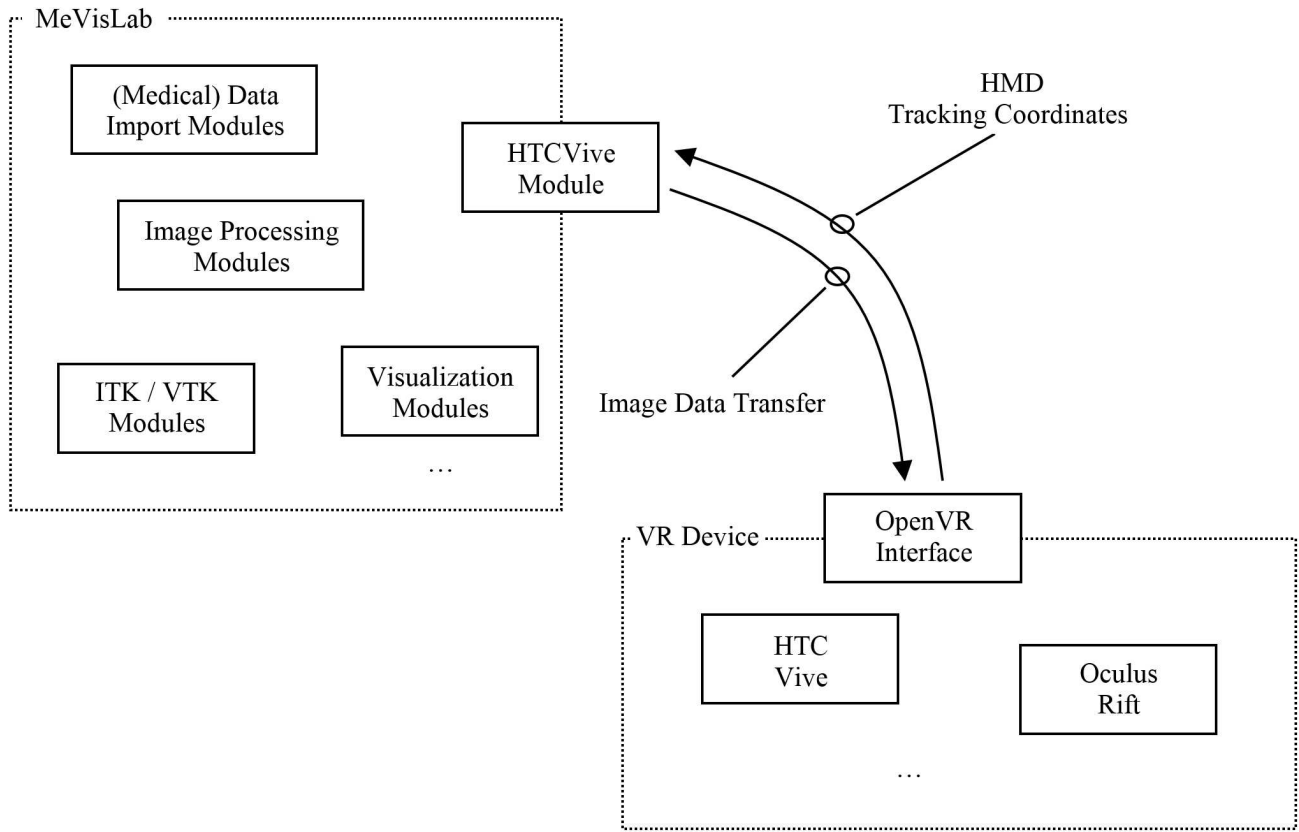

**Fig 5. High level workflow diagram that demonstrates the communication and interaction between the MeVisLab platform and the HTC Vive device via the OpenVR library.**

https://doi.org/10.1371/journal.pone.0173972.g005

wearing the HTC Vive headset sees. Note that the HMD tracking coordinates are transferred via direct parameter connections in our example. The remaining modules in the network example from Fig 6 are mainly arithmetic and matrices modules to decompose (DecomposeMatrix and DecomposeMatrix2) and compose (ComposeMatrix) the rotation and translation for a correct visualization. The interface and parameters of our HTCVive module are shown on the left side (Panel HTCVive). Finally, a status window of the SteamVR is presented on the lower right corner (black), indicating that the HMD, the two HTC Vive controllers and the two HTC Vive Lighthouse base stations are ready and running (green).

## Results

The aim of this contribution was to investigate the feasibility of using the Virtual Reality device HTC Vive under the medical prototyping platform MeVisLab. We present the direct rendering and visualization of (medical) image data in the head mounted device using the publicly available OpenVR library. The software integration was achieved under Microsoft Windows 8.1 with MeVisLab version 2.8.1 (21-06-2016) for Visual Studio 2015 x64 (http://www.mevislab.de/download/) and the OpenVR SDK version 1.0.2 (https://github.com/ValveSoftware/openvr). The MeVisLab module was implemented as an image processing module in C++ using the MeVisLab Project Wizard and the Microsoft Visual Studio 2015 Community Edition. The HTC Vive MeVisLab module (HTCVive) has an input for the medical image data, which is transferred and displayed in the HTC Vive. Additionally, the HTCVive module provides the data at

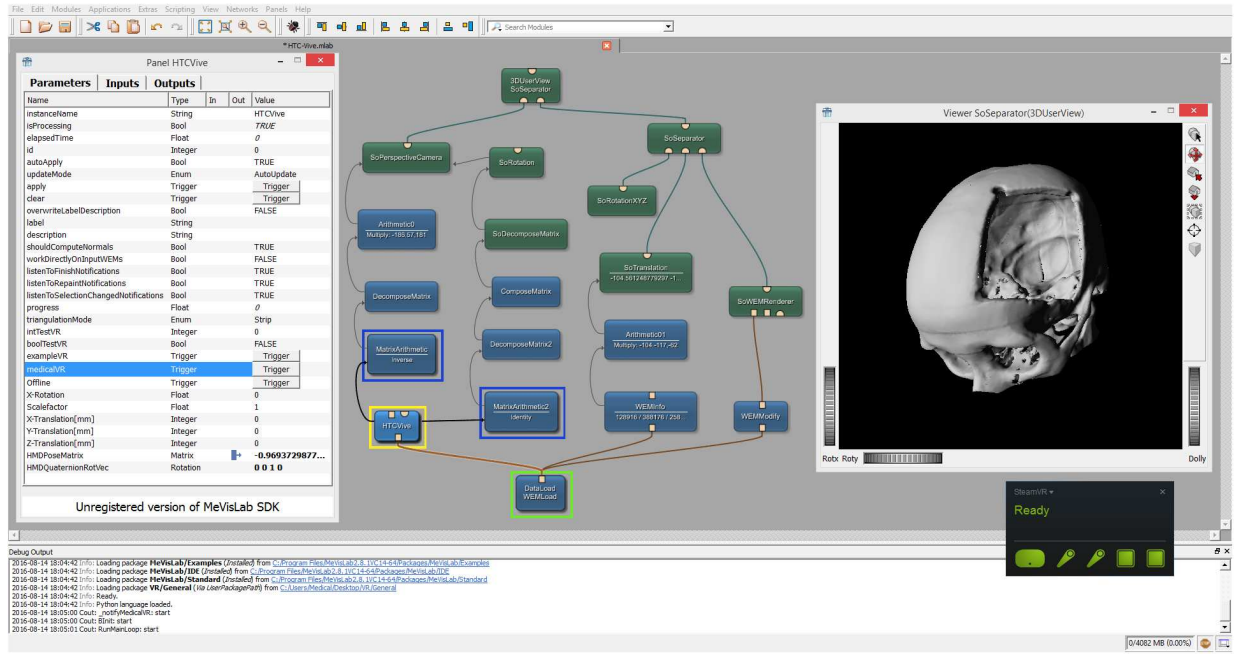

**Fig 6. Screenshot of the comprehensive MeVisLab network with the HTCVive module (blue, lower left corner surrounded by a yellow rectangle) and its interface and parameters on the left side (Panel HTCVive).** Additionally, the module provides the HMD tracking coordinates of the HTC Vive, which can be used to have the same position and viewing direction as the actual user wearing the HTC Vive in an extra 3D view (Viewer SoSeparator(3DUserView) on the right side).

https://doi.org/10.1371/journal.pone.0173972.g006

an output (rectangle) for visualization in a standard undistorted 3D view, and the module provides the HMD tracking coordinates of the HTC Vive (Fig 6, right window). Finally, an extra window can be created for a distorted view of the virtual reality input for the left and the right eye (Fig 7 without texture and Fig 8 with texture). For an evaluation, we tested several medical datasets, like patient skulls with cranial defects and scans from gastrointestinal tracts for usage in virtual colonoscopy (Fig 9). As a hardware setting, we applied two computers: a desktop PC and a MacBook Pro laptop, which both had Windows 8.1 Enterprise installed. The hardware

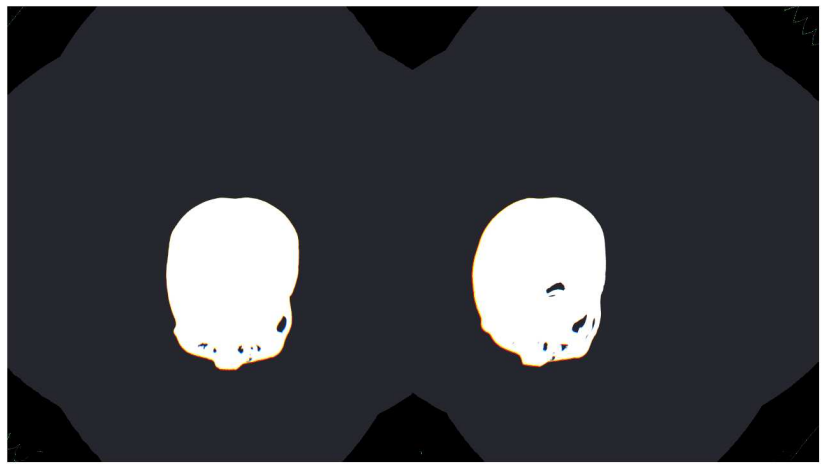

**Fig 7. The distorted view of the VR input for the right and left eye without texture.**

https://doi.org/10.1371/journal.pone.0173972.g007

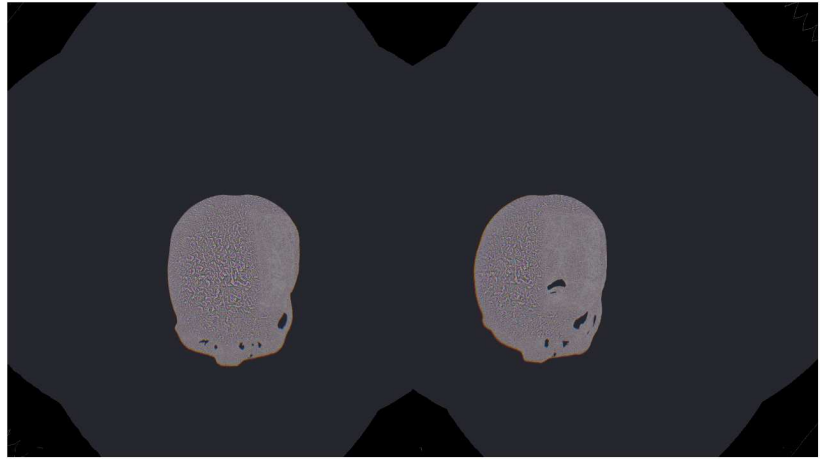

**Fig 8. The distorted view of the VR input for the right and left eye with texture.**

https://doi.org/10.1371/journal.pone.0173972.g008

**Fig 9. The figure shows a scan of the gastrointestinal tract that has been imported and displayed in the HTC Vive for usage in virtual colonoscopy.**

https://doi.org/10.1371/journal.pone.0173972.g009

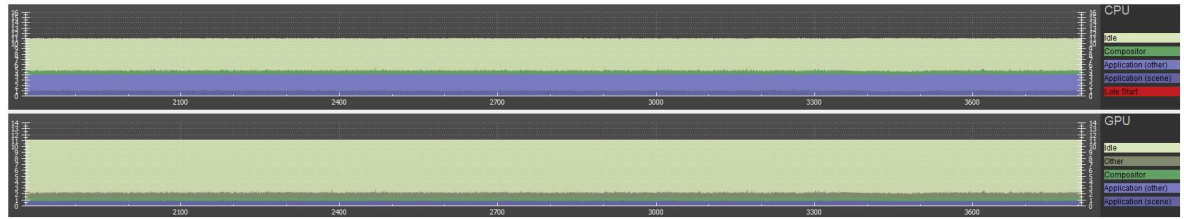

**Fig 10. Frame timing result diagrams of the CPU (top) and the GPU (bottom) for a Windows desktop PC with Intel Core i7-3770 CPU @ 3.40GHz, 16 GB RAM and a NVIDIA GeForce GTX 970 graphic card.**

https://doi.org/10.1371/journal.pone.0173972.g010

configuration of the desktop PC consisted of an Intel Core i7-3770 CPU @ 3.40GHz, 16 GB RAM and a NVIDIA GeForce GTX 970 graphics card, and the hardware configuration of the laptop computer consisted of an Intel Core i7-4850HQ CPU @ 2.30GHz, 16 GB RAM and a NVIDIA GeForce GT 750M graphics card. Figs 10 and 11 show the frame timing results of both configurations for the CPU (top) and the GPU (bottom). The frame timing results come directly from the SteamVR and can be activated and displayed via the SteamVR status window. As seen, the default view splits the CPU and GPU performance in a pair of stacked graphs: The blue sections represent the time spent by the application, which is further spitted between application-scene and application-other. Application-scene is the amount of work performed in between, when WaitGetPoses (returns pose(s) to use to render scene) returns and the second eye texture is submitted. Application-other is any time spent after this for rendering the application's companion window, etc. Note that the CPU timing does not capture any parallel work being performed, for example, on the application's main thread. The brown section (other) in the GPU timing reflects any GPU bubbles, context switching overhead, or GPU work from other application getting scheduled in between other segments of work for that frame (to examine this in more detail requires the use of gpuview). For more information, please see also (last access January 2017):

https://developer.valvesoftware.com/wiki/SteamVR/Frame_Timing.

As shown in the frame timing diagram in Fig 10, a sufficient framerate for the desktop PC configuration could be achieved, because ten frames per second are already considered as real-time or interactive in computer graphics applications [20]. Hence, the medical datasets could be displayed very pleasant to the human eye inside the HTC Vive device. In contrast, the framerates on the laptop were in general not sufficient, resulting in a flickering inside the HTC Vive. For a short period of time, this is acceptable as a temporary (mobile) solution, but for a longer working period, this is currently definitely not suitable. However, the laptop and its

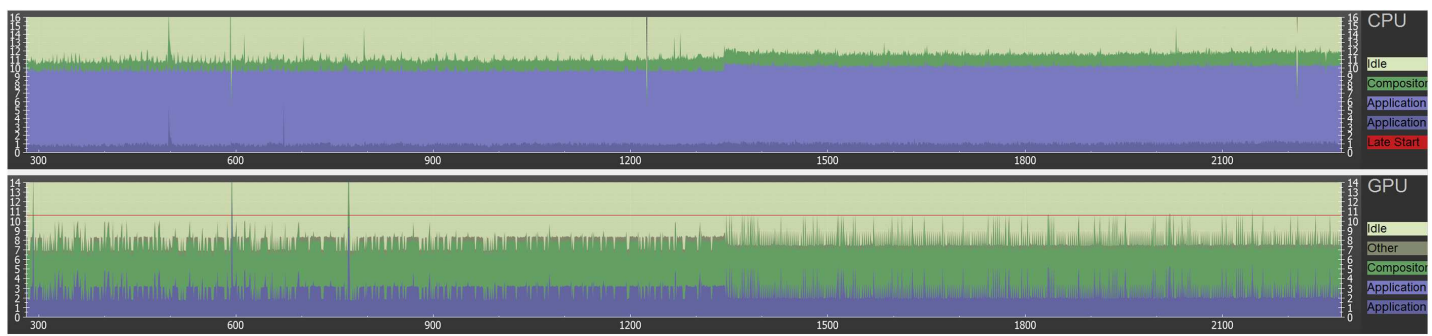

**Fig 11. Frame timing result diagrams of the CPU (top) and the GPU (bottom) for a Windows laptop with Intel Core i7-4850HQ CPU @ 2.30GHz, 16 GB RAM and a NVIDIA GeForce GT 750M graphic card.**

https://doi.org/10.1371/journal.pone.0173972.g011

hardware is also not considered as HTC Vive ready, but we expect to see soon laptops that can handle the HTC Vive requirements, so it can also be used as a portable application.

## Discussion

In this open access article, we demonstrated the successful integration and evaluation of the HTC Vive headset into the research prototyping platform MeVisLab via the OpenVR library. OpenVR is a software development kit and application programming interface developed by Valve (http://www.valvesoftware.com/) for supporting the SteamVR (HTC Vive) and other VR headsets. Hence, our integration is device and vendor independent and can also be used with other devices. The proposed software integration has been carried out with the C++ implementation of a MeVisLab image processing module that provides inputs for medical datasets in different formats. Thus, our module is easy to use and can be added to existing MeVisLab networks or applied in new ones. Per drag and drop, data outputs under MeVisLab can be directly connected to an input of our module to transfer the data into the HTC Vive headset and display it there. Furthermore, dataset operations and manipulations, like translation, rotation, segmentations, etc. that have been performed in advance with corresponding MeVisLab modules will also directly be shown inside the headset. Highlights of the proposed contribution are as follows:

- Successful integration of OpenVR into MeVisLab

- Solution enables MeVisLab networks to connect to VR headsets

- Real-time visualization of medical data in VR under MeVisLab

- Proof of concept evaluation with the HTC Vive headset

- HTC Vive MeVisLab module can be added to existing MeVisLab networks.

Areas of future work include the evaluation of our integration with a greater amount of medical data and formats. Although if we did not encounter motion sickness in our tests, this might be an issue for some users and needs to be explored systematically. In addition, we plan a MeVisLab module communicating with the Oculus Rift head mounted display, or, as an alternative, the extension of our VR module supporting also the Oculus Rift, when fully supported by the OpenVR library, and the evaluation on non-Windows-Setups like Linux or Mac OS. Moreover, we work on applying our solution to support computer-aided reconstructions of facial defects [21], [22] with photorealistic rendering in VR [23]. This will enable an even more realistic assessment of pre-operative planning results. Furthermore, we investigate options for performing image-guided therapy tasks inside VR, for example, the planning of facial interventions using so called miniplates [24] and virtual stent simulations to treat aneurysms [25–29]. Capturing and reconstruction of 3D models of tumors in the gastrointestinal tract during diagnostic endoscopic procedures will ease the planning for the endoscopic removal without the need for prolonged or repeated anesthesia [30]. Using segmentation algorithms implemented in MeVisLab [31–33] will help identify areas of tumor spread in the 3D model due to surface characteristics. Interactive segmentation [34], [35] might even be done with the HTC Vive controllers. Additionally, reviewing 3D models of tumors in virtual space will provide a platform for training to improve recognition of these lesions from different angles. We would like to pursue the development of an Augmented Reality (AR) module for MeVisLab, to support novel devices like the Microsoft HoloLens, because optical see-through head-mounted displays seem to be very promising during surgical navigation [36]. Last but not least, we plan to integrate OpenVR into other research platforms, like OsiriX [37], 3D

Slicer [38] or the two Medical Imaging (Interaction) Toolkits (MITK) from Beijing in China (www.mitk.net) [39] and Heidelberg in Germany (www.mitk.org) [40], respectively.

## Acknowledgments

The work received funding from BioTechMed-Graz in Austria (https://biotechmedgraz.at/en/, "Hardware accelerated intelligent medical imaging"), the 6[th] Call of the Initial Funding Program from the Research & Technology House (F&T-Haus) at the Graz University of Technology (https://www.tugraz.at/en/, "Interactive Planning and Reconstruction of Facial Defects", PI: Jan Egger) and was supported by TU Graz Open Access Publishing Fund. Dr. Xiaojun Chen receives support by the Natural Science Foundation of China (www.nsfc.gov.cn, Grant No.: 81511130089) and the Foundation of Science and Technology Commission of Shanghai Municipality (Grants No.: 14441901002, 15510722200 and 16441908400). Moreover, the authors want to thank their colleagues Laurenz Theuerkauf, Peter Mohr and Denis Kalkofen for their support. Furthermore, we are planning to provide the module as Open Source to the research community in the future. Until then, the module can be requested from the authors. Finally, a video demonstrating the integration can be found on the following YouTube channel:

https://www.youtube.com/c/JanEgger

## Author Contributions

**Conceptualization:** JE MG.

**Data curation:** JW XC XL.

**Formal analysis:** JE MG.

**Funding acquisition:** JE DS XC.

**Investigation:** JE MG.

**Methodology:** JE MG.

**Project administration:** JE DS.

**Resources:** JE MG JW PB AH XL XC DS.

**Software:** JE MG PB.

**Supervision:** JE DS.

**Validation:** JE MG.

**Visualization:** JE MG.

**Writing – original draft:** JE.

**Writing – review & editing:** JE DS.